\documentclass[fp,twocolumn]{jpsj3}
\usepackage{txfonts}
\usepackage{graphicx}
\usepackage{bm}
\usepackage{braket}
\usepackage{mathtools}
\usepackage{color}

\title{Higher-Order Topological Phase in a Honeycomb-Lattice Model\\ 
with Anti-Kekul\'{e} Distortion}

\author{Tomonari Mizoguchi, Hiromu Araki, and Yasuhiro Hatsugai}
\inst{Department of Physics, University of Tsukuba, Tsukuba, Ibaraki 305-8571, Japan}

\abst{
Higher-order topological insulators have attracted considerable interests 
as a novel topological phase of matter, where topologically non-trivial nature of bulk 
protects boundary states whose co-dimension is larger than one.  
It has been revealed that the alternating pattern of hopping amplitudes 
in two-dimensional lattices provides a promising 
route to realization of the higher-order topological insulators.
In this paper, we propose that a honeycomb-lattice model with anti-Kekul\'{e} distortion hosts 
a higher-order topological phase. 
Here, the term anti-Kekul\'{e} distortion means that the pattern of strong and weak hoppings 
is opposite to that for the conventional Kekul\'{e} distortion.
We demonstrate the existence of the higher-order topological phase
by calculating the $\mathbb{Z}_6$ Berry phase that serves as a bulk topological invariant of the higher-order topological phase, 
and by showing the existence of corner states. 
}

\begin{document}
\maketitle
\section{Introduction}
A honeycomb lattice provides us of a platform to 
realize intriguing phenomena in solid-state physics.
It serves as one of the simplest examples of a non-Bravais lattice, 
which gives rise to a multiband electronic structure even for a single-orbital, nearest-neighbor (NN) tight-binding model. 
A typical example of the honeycomb-lattice systems is graphene~\cite{CastroNeto2009}, 
where emergence of Dirac fermions in solid has attracted considerable interests for decades.
Besides solid-state systems, the implementation of the honeycomb structure 
in mechanical systems~\cite{Cserti2004,Torrent2013,Kariyado2015} and photonic crystals~\cite{Chern2003,Lu2014} has also been actively studied. 

\begin{figure}[b]
\begin{center}
\includegraphics[width=0.85\linewidth]{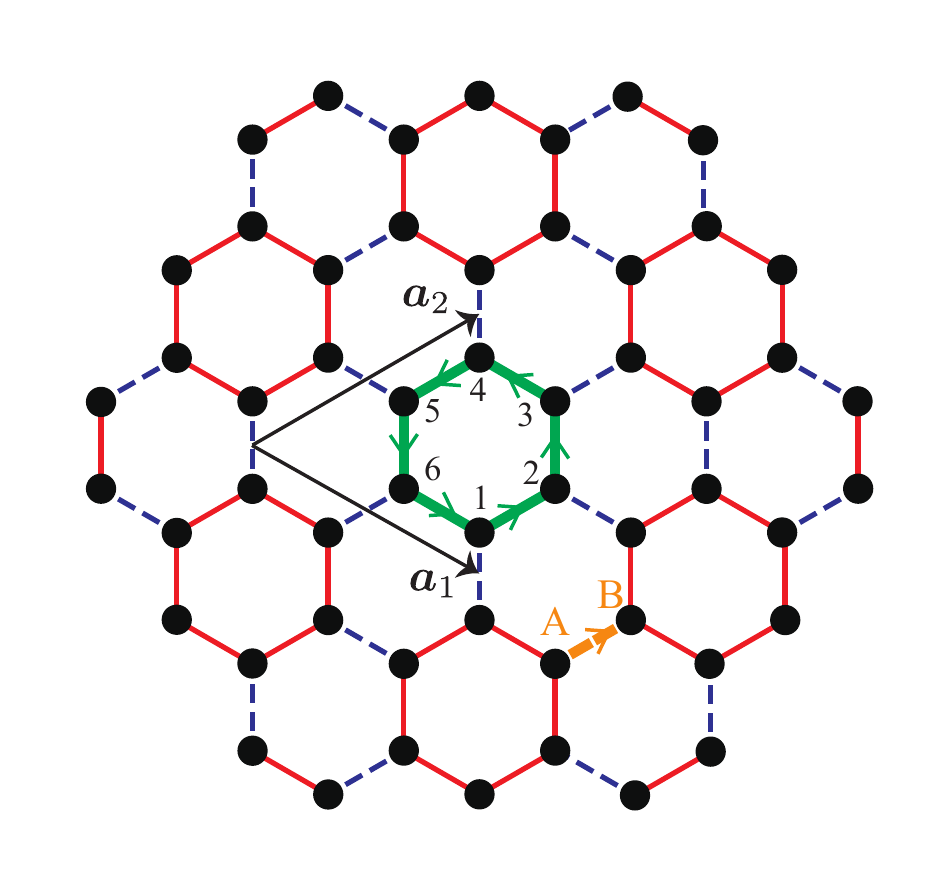}
\vspace{-10pt}
\caption{(Color online) A honeycomb-lattice model considered in paper.
The solid (dashed) bonds have a hopping $t_0$ ($t_1$). 
The lattice vectors $\bm{a}_1$ and $\bm{a}_2$ and the sublattices 1-6 are indicated in the figure.
The twist in $H(\bm{\Theta})$ is introduced at a green bold hexagon,
and the twist in $H(\theta)$ is introduced at an orange bold bond
(see Sect.~\ref{sec:berry}).}
  \label{fig:1}
 \end{center}
\end{figure}

Spatial modulation on a honeycomb lattice makes its physics more intriguing. 
One of the typical modulation patterns is Kekul\'{e}-type modulation~\cite{Ajiki1994,Chamon2000,Hou2007,Chamon2008,Frank2011,Wu2012}
(Fig.~\ref{fig:1}),
which results in the enlargement of the unit cell. 
Recently, such a modulated structure has been revisited in the context of 
topological phases protected by crystalline symmetries~\cite{Koshino2014,Wu2016,Kariyado2017,Liu2019,Lee2019,ZangenehNejad2019}. 
Indeed, it has been found that the single-orbital tight-binding model 
hosts a topological crystalline insulator, accompanied by characteristic edge states~\cite{Wu2016,Kariyado2017}
as a typical example of the bulk-edge correspondence~\cite{Hatsugai1993}. 
Such an implementation of the topological nature is also applied to photonic crystals~\cite{Wu2015,Yang2018,Li2018}.

In this paper, 
we present a different view of topological phases in 
the honeycomb-lattice model with the Kekul\'{e}-type modulation.
More precisely, we propose that the present model hosts a higher-order topological phase
not only for the Kekul\'{e} modulation but also for the anti-Kekul\'{e} modulation.
Throughout this paper, 
the Kekul\'{e} modulation denotes the case where the intra-hexagonal hopping is smaller than the other hoppings,
and the anti-Kekul\'{e} modulation denotes the opposite case; see Sect.~\ref{sec:model} for details.

Higher-order topological insulators (HOTIs)
are a novel topological phase of matter which exhibit unconventional bulk-boundary correspondence.
Namely, in HOTIs, the characteristic boundary states appear at the boundary with co-dimension larger than one~\cite{Benalcazar2017,Song2017,Schindler2018,Ezawa2018,Xu2017,Kunst2018,Fukui2018}. 
Several tight-binding models have been proposed for the HOTI in two dimensions,
such as the Benalcazar-Bernevig-Hughes (BBH) model and the breathing kagome model~\cite{Ezawa2018,Xu2017,Kunst2018}.
These models can be regarded as two-dimensional analogs of the Su-Schriffer-Heeger (SSH) model~\cite{Su1979},
in a sense that the alternating structures of strong and weak hoppings play an essential role in 
the gap-opening of bulk bands and in the emergence of corner states.
Considering these, a honeycomb-lattice model with the Kekul\'{e}-type modulation is also a promising platform for the HOTI.
Indeed, the HOTI was found for the Kekul\'{e} modulation~\cite{Liu2019,Lee2019,ZangenehNejad2019}.
Here we propose that the case for the anti-Kekul\'{e} modulation also hosts a HOTI,
which is different from that for the Kekul\'{e} modulation.

To capture the higher-order topology of the present model with anti-Kekul\'{e} distortion, 
we show the topological invariant that characterized the HOTI phase.
Very recently, the authors have proposed that the HOTIs are characterized by a bulk topological invariant, 
namely, the $\mathbb{Z}_Q$ Berry phase defined with respect to the 
parameters of the local twist of the Hamiltonian~\cite{Araki2019}. 
By using this topological invariant, 
we demonstrate that 
the $\mathbb{Z}_Q$ Berry phase 
is quantized in six-fold (i.e., we obtain the $\mathbb{Z}_6$ Berry phase as a topological invariant)
due to the $C_{6v}$ point group symmetry which contains
the six-fold symmetry around a certain hexagonal plaquette,
and that 
the $\mathbb{Z}_6$ Berry phase is $\pi$ for anti-Kekul\'{e} modulation, while it is zero 
for the Kekul\'{e} modulation. 
We then show the existence of corner states when the $\mathbb{Z}_6$ Berry phase is $\pi$.
For completeness, we also show that 
the different Berry phase, quantized in $\mathbb{Z}_2$, can be defined as well, 
and that this Berry phase characterizes the HOTI phase for Kekul\'{e} modulation.

The remainder of this paper is organized as follows. 
In Sect.~\ref{sec:model}, we introduce the model considered in this paper, and summarize the characters of band structures in a bulk. 
In Sect.~\ref{sec:berry}, 
we perform the 
Berry-phase analyses, and show that topologically nontrivial phases 
exist in for both Kekul\'{e} and anti-Kekul\'{e} modulations.
Then, to demonstrate that the non-trivial Berry phase 
results in the emergence of the topological boundary states,
we calculate the energy spectra and the wave functions 
of finite samples under the open boundary conditions 
in Sect.~\ref{sec:open}.
Section~\ref{sec:1/6} is devoted to the 
analysis of a $1/6$-filled system, where we also find the quantization of 
the $\mathbb{Z}_6$ Berry phase.
Finally, in Sect.~\ref{sec:summary}, we summarize 
this paper and discuss future perspectives.

\section{Model \label{sec:model}}
\begin{figure}[b]
\begin{center}
\includegraphics[width=0.95\linewidth]{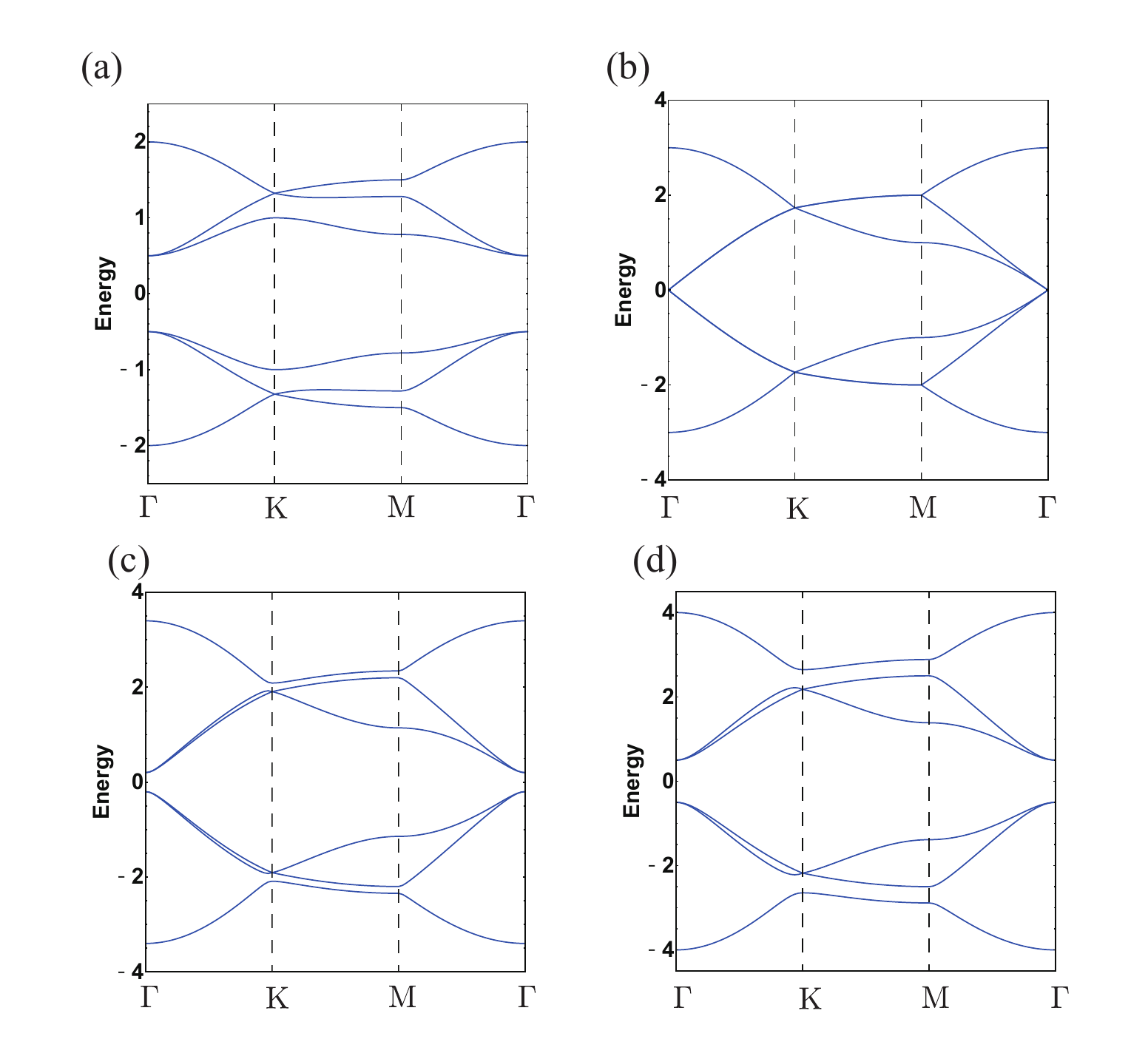}
\vspace{-10pt}
\caption{(Color online) Band structures for $(t_0,t_1) = $ (a) $(0.5,1)$, 
(b) $(1,1)$, (c) $(1.2,1)$, and (d) $(1.5,1)$.
$\Gamma$, K, and M are high-symmetry points in the Brillouin zone. 
}
  \label{fig:2}
 \end{center}
\end{figure}
We consider the Hamiltonian on a honeycomb lattice:
\begin{eqnarray}
H = \sum_{\langle i,j\rangle} t_{i,j} c^{\dagger}_i c_j + (\mathrm{h.c.}), \label{eq:Hamiltonian}
\end{eqnarray}
where $\langle i,j \rangle$ represents the NN pairs on the honeycomb lattice,
and the transfer integral $t_{i,j}$ is equal to $t_0$ ($t_1$) on solid (dashed) bonds
in Fig.~\ref{fig:1}(a). 
The parameter region $t_0 < t_1$ ($t_0 > t_1$) corresponds to 
the Kekul\'{e} (anti-Kekul\'{e}) modulation.
We remark that the present model is different from that considered in Ref.~\cite{Ezawa2018_2},
which also studied the HOTI on a honeycomb lattice. 
In this paper, if not stated otherwise, we consider the case of half-filling.

The Hamiltonian of Eq. (\ref{eq:Hamiltonian}) 
can be written in the momentum-space representation as
\begin{eqnarray}
H =\sum_{\bm{k}} \bm{\psi}_{\bm{k}}^\dagger \mathcal{H}_{\bm{k}} \bm{\psi}_{\bm{k}},
\end{eqnarray}
where $\bm{k} = (k_x,k_y)$, 
$\bm{\psi}_{\bm{k}} = (c_{\bm{k},1},c_{\bm{k},2},c_{\bm{k},3},c_{\bm{k},4},c_{\bm{k},5},c_{\bm{k},6})^{\mathrm{T}}$,
and
\begin{eqnarray*}
\scriptsize
\mathcal{H}_{\bm{k}}=   
\left(
\begin{array}{cccccc}
0 & t_0 & 0 & t_1e^{i \bm{k} \cdot (\bm{a}_{1}-\bm{a}_2) } & 0 &  t_0  \\
t_0 & 0 & t_0 & 0 & t_1 e^{i \bm{k} \cdot \bm{a}_1} & 0 \\
0 & t_0 & 0 & t_0 & 0 & t_1 e^{i \bm{k} \cdot \bm{a}_2} \\
t_1e^{- i \bm{k} \cdot (\bm{a}_{1}-\bm{a}_2) } & 0 & t_0 & 0 & t_0 & 0 \\
0 & t_1e^{-i \bm{k} \cdot \bm{a}_1} & 0 & t_0 & 0 & t_0 \\
t_0 & 0 & t_1e^{-i \bm{k} \cdot \bm{a}_2} & 0 & t_0 & 0 \\
\end{array}
\right). \nonumber \\
\end{eqnarray*}
For the sublattices and the lattice vectors, see Fig.~\ref{fig:1}.
The Hamiltonian has a chiral symmetry expressed by a matrix
\begin{eqnarray}
\gamma = \left(
\begin{array}{cccccc}
1 & 0&0 &0 &0 &0 \\
 0& -1 & 0& 0& 0&0 \\  
 0& 0& 1 & 0& 0&0 \\
 0& 0& 0& -1 & 0&0 \\
0 & 0& 0& 0& 1 &0 \\
0 &0 &0 &0 &0 &  -1 \\ 
 \end{array}
\right),
\end{eqnarray}
which satisfies $\gamma^2 = 1$ and transforms $\mathcal{H}_{\bm{k}}$ as 
\begin{eqnarray}
\gamma \mathcal{H}_{\bm{k}} \gamma = - \mathcal{H}_{\bm{k}}.
\end{eqnarray}
This means that the positive eigenvalues and negative eigenvalues always appear in a pairwise manner.
The model also has a six-fold rotational symmetry,
\begin{eqnarray}
U^{\dagger}_{C_6} \mathcal{H}_{\bm{k}}U_{C_6} =  \mathcal{H}_{C_6(\bm{k})},
\end{eqnarray}
with 
\begin{eqnarray}
U_{C_6} = \left(
\begin{array}{cccccc}
 0& 0&0 &0 &0 & 1 \\
 1& 0 &0 & 0& 0&0 \\  
0 &1 & 0 & 0&0 &0 \\
0 &0 & 1& 0 & 0& 0\\
0 &0 & 0& 1&0 &0 \\
0 &0 & 0& 0&1 & 0 \\ 
 \end{array}
\right),
\end{eqnarray}
and $C_6(\bm{k}) = \left( \frac{k_x -\sqrt{3}k_y}{2}, \frac{\sqrt{3} k_x +  k_y}{2}\right)$.

In Fig.~\ref{fig:2} we show the bulk band structures for several parameters.
The energy gap of the ground state for the half-filled system 
corresponds to the band gap between the third and the fourth band, and there is a finite band gap when $t_0 \neq t_1$. 
At $t_0 =t_1$, the Hamiltonian is reduced to the conventional 
NN hopping model on a honeycomb lattice, and thus the band gap closes.
This follows from the analytical expression of the energy eigenvalues at $\Gamma$ points,
namely, $\varepsilon = \pm (2 t_0 +t_1)$, $\pm (t_0 -t_1)$. Note that the eigenvalues $\pm (t_0 -t_1)$ have two-fold degeneracy, as shown in Fig.~\ref{fig:2}.
For the discussion in Sect.~\ref{sec:1/6}, where we study the case of $1/6$-filling, 
we also mention the band gap between the lowest and the second bands. 
We see that, for $t_0 < t_1$, the bands touches at K point. 
For $t_0 > t_1$, the touching disappears. 
However, the indirect band gap (i.e., the energy difference between two bands at different momenta) is negative
for $1 < t_0/ t_1 \lesssim 1.32$, 
meaning that the system with $1/6$-filling is semimetallic in this region.
For $t_0/ t_1 \gtrsim 1.32$, both direct and indirect band gaps are positive, thus the system with $1/6$-filling is insulating.

Topological physics of the present model has been studied in various contexts. For instance, the previous study by Kariyado and Hu~\cite{Kariyado2017} has elucidated that 
the mirror winding number characterizes the topological phases.
In fact, the mirror winding number depends on the choice of the unit cell, 
which coincides with the fact that emergence of the edge modes 
depends on the shape of the edges. 
The mirror winding number changes at the critical point, 
$t_0 =t_1$. 
Recently, the HOTI phase of the model was also studied~\cite{Liu2019,Lee2019,ZangenehNejad2019}.
The previous works focused on the case with $t_0 < t_1$, i.e., the Kekul\'{e} modulation.
The existence of the HOTI phase in this region was demonstrated 
by using the quadrupole moment~\cite{Liu2019}, the Wilson loop and the nested Wilson loop~\cite{Lee2019}, and the bulk poralization~\cite{ZangenehNejad2019}. 
In the present paper, we treat both Kekul\'{e} $(t_0 < t_1)$ and anti-Kekul\'{e} $(t_0 > t_1)$ modulations,
and show the existence of HOTI phases
by using two different Berry phases, namely, the $\mathbb{Z}_2$ Berry phase (for the Kekul\'{e} modulation) and 
the $\mathbb{Z}_6$ Berry phase (for the anti-Kekul\'{e} modulation), as we will discuss in the next section.

\section{Berry phases and higher-order topology \label{sec:berry}}
\begin{figure}[b]
\begin{center}
\includegraphics[width=.95\linewidth]{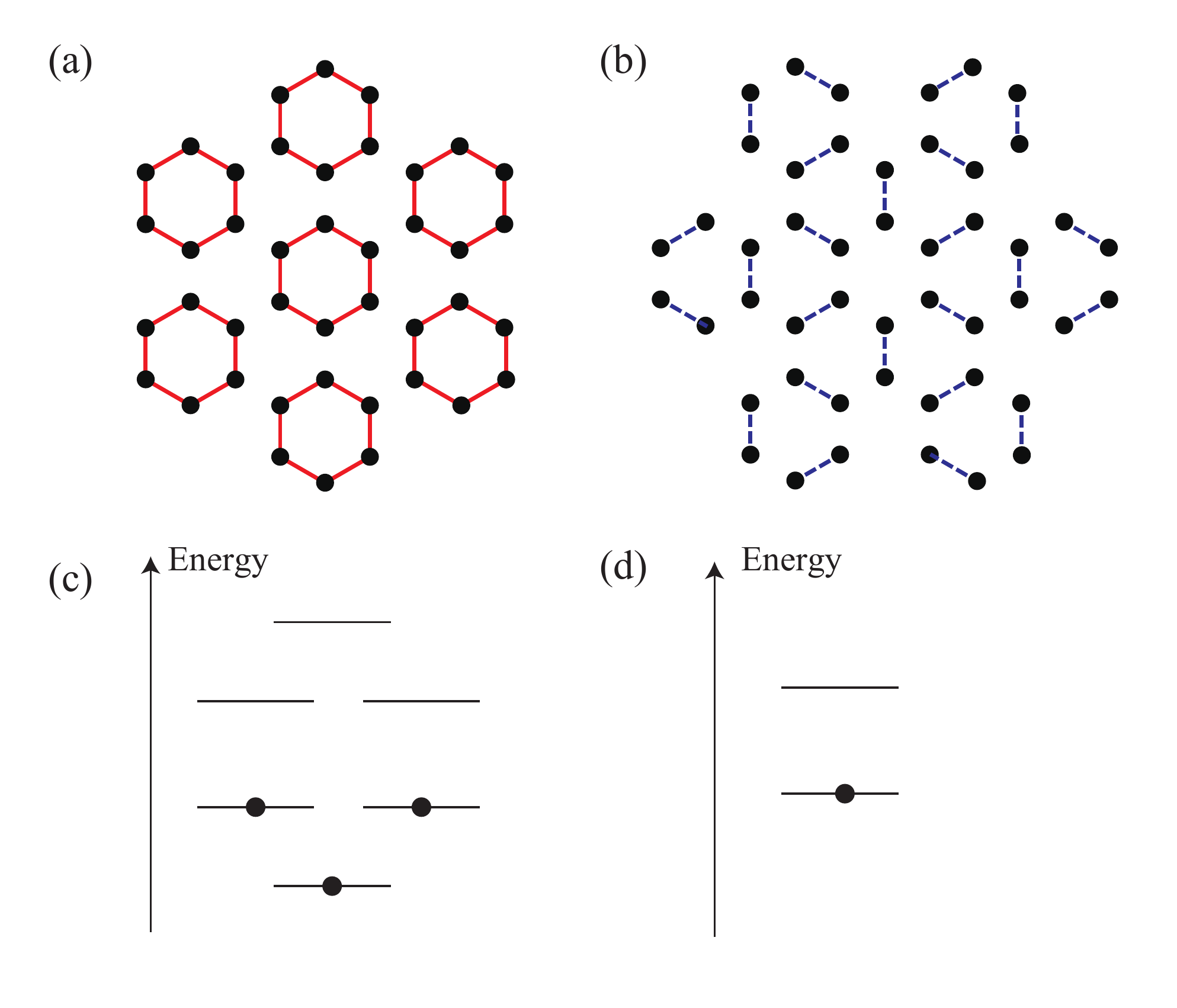}
\vspace{-10pt}
\caption{(Color online) Schematic figures of (a) the hexamer state connected to the ground state for $t_0 > t_1$, 
and (b) the dimer state connected to the ground state for $t_0 < t_1$.
The schematic energy diagrams for (c) an isolated hexamer 
and (d)  for an isolated dimer. 
Black dots mean that the state is occupied. }
  \label{fig:3}
 \end{center}
\end{figure}
In this section, we investigate the topological property through 
the analysis of the $\mathbb{Z}_Q$ Berry phase.
Note that the $\mathbb{Z}_Q$ Berry phase does not depend on the choice of the unit cell, as we will show below.
We also remark that the $\mathbb{Z}_Q$ Berry phase is defined for the system on a torus, i.e., under the periodic boundary conditions in both of two directions.

Before going into the detailed analysis, 
it is instructive to see what the implication of the non-trivial Berry phase is~\cite{Kariyado2014}.
In fact, physical pictures of the ground states in this model can be well-illustrated by considering the 
two limiting cases, i.e., (i) $t_0 \neq 0$, $t_1=0$ and (ii) $t_0 = 0$, $t_1 \neq 0$. 
In case (i), the model is reduced to a set of isolated hexagons, as illustrated in Fig.~\ref{fig:3}(a), 
while it is reduced to a set of isolated dimers for case (ii), as illustrated in Fig.~\ref{fig:3}(b).
The ground states in these limits are product states where each isolated unit is half-filled.
Clearly, both of these states are gapped at the half-filling [Fig.~\ref{fig:3}(c) and \ref{fig:3}(d)].
Further, both of these states are topologically nontrivial in a sense that they cannot be decomposed into the product state of individual atomic states.
Now, let us consider 
the case where both $t_0$ and $t_1$ are non-zero.
Since the energy gap is not closed for $t_0 \neq t_1$,
the ground state is adiabatically connected to either of
the irreducible cluster states, except for the critical point $t_0 = t_1$. 
This can be evidenced by investigating the topological invariant;
if we properly define the topological invariant, or the $\mathbb{Z}_Q$ Berry phase in this case, such that 
its value is non-trivial in the limit of the irreducible cluster state, 
the non-trivial topological invariant of the ground state 
for a certain parameter directly means the
adiabatic connection to the irreducible cluster state,
because the topological invariant, from its definition, does not change unless the gap is closed.
It is worth noting that the hexamer state and the dimer state are captured by the different topological invariants.
To be specific, the hexamer state can be captured by the $\mathbb{Z}_6$ Berry phase,
while the dimer state by the $\mathbb{Z}_2$ Berry phase, as we will elucidate in the following.

\subsection{$\mathbb{Z}_6$ Berry phase}
\begin{figure}[t]
\begin{center}
\includegraphics[width=0.95\linewidth]{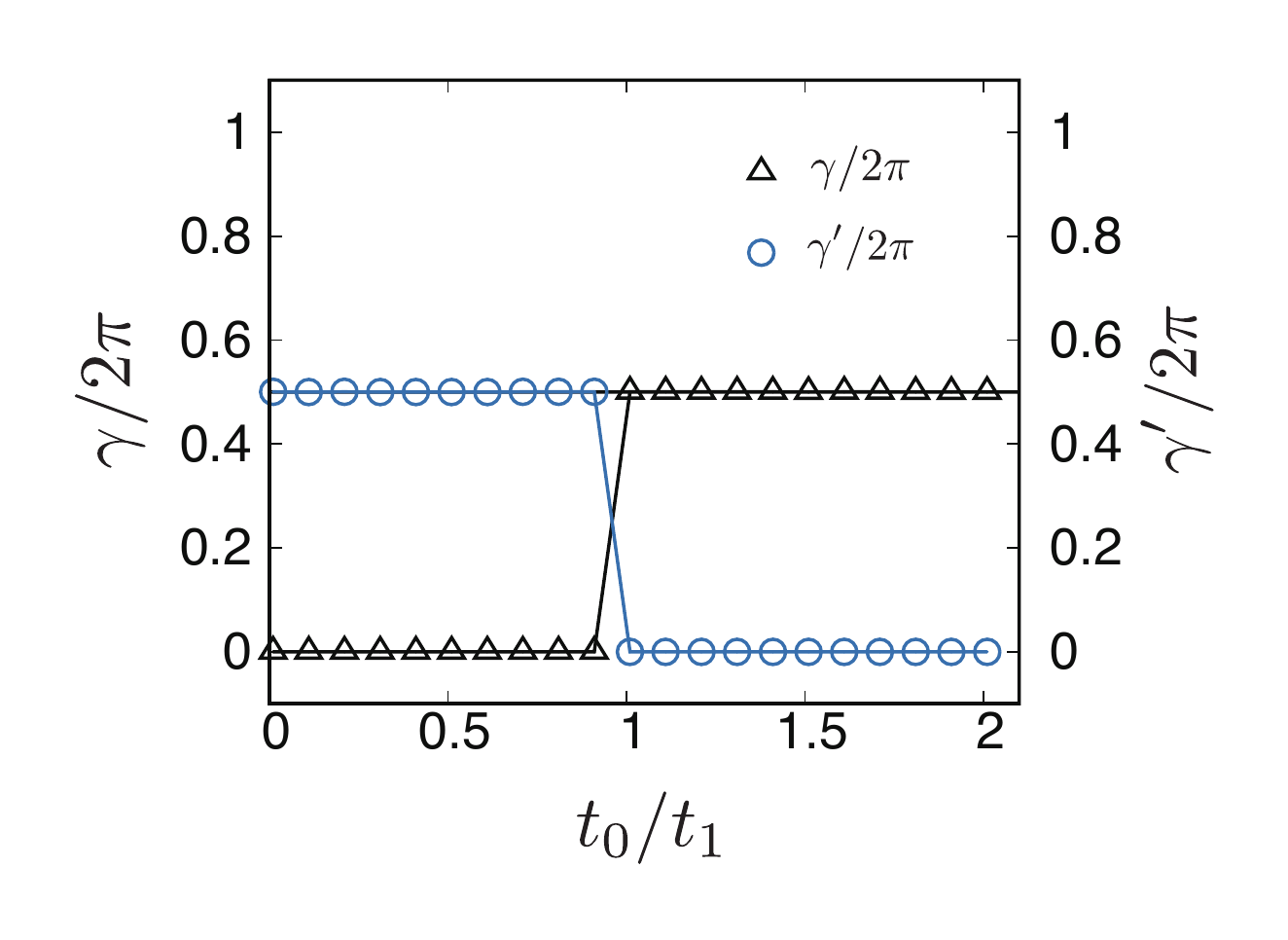}
\vspace{-10pt}
\caption{(Color online)
The $\mathbb{Z}_6$ Berry phase (black open triangles)
and the $\mathbb{Z}_2$ Berry phase (blue open circles)
as a function of $t_0 /t_1$.
The numerical calculation is performed on a 576-site system.
The data for $t_0/t_1 = 0.1 n + 0.01$ ($n= 0, \cdots 20$), 
and a solid lines are for the guide to the eyes.}
  \label{fig:4}
 \end{center}
\end{figure}
We first calculate the Berry phase that captures the hexagonal cluster.
Such a Berry phase can be defined by introducing the local twist of the Hamiltonian 
on a certain hexagonal plaquette~\cite{Araki2019,Hatsugai2011}.
To be concrete, we first pick up one of the hexagonel plaquettes all of whose bonds 
have the hopping $t_0$ [see a green bold hexagon in Fig.~\ref{fig:1}]. 
We write the Hamiltonian on this plaquette as $h_0$, and the Hamiltonian 
can be divided in two parts, as $H = h_0 + (H-h_0)$.  
We then introduce the twist of the Hamiltonian 
only in $h_0$, while keeping $(H-h_0)$ unchanged.
To be specific, we consider the twisted Hamiltonian
$h_0 (\bm{\Theta})$, where
$\bm{\Theta} = \left( \theta_1,\theta_2, \theta_3, \theta_4, \theta_5 \right)$
with $\theta_i \in [0,2\pi]$ are parameters describing the twist angles.
The explicit form of $h_0 (\bm{\Theta})$ is 
\begin{eqnarray}
h_0(\bm{\Theta}) = t_0 \sum_{a= 1}^{6} 
e^{i\theta_{a+1}}c^{\dagger}_{a+1} c_{a} + (\mathrm{h.c.}),
\end{eqnarray}
where we have introduced $\theta_6 := -(\theta_1 + \theta_2 + \theta_3 + \theta_4 + \theta_5)$
and have defined $a$ in modulo $6$ such that $a+6 \equiv a$; 
the labels of sites 1-6 follow those for the sublattices shown in Fig.~\ref{fig:1}(a).
The total Hamiltonian with the twist is given as 
$H(\bm{\Theta}) = h_0(\bm{\Theta}) + (H-h_0)$.
Then, for the many-body ground state of $H (\bm{\Theta})$, which we write as $\ket{\Psi (\bm{\Theta}) }$, 
we define the Berry connection as 
\begin{eqnarray}
\bm{A} (\bm{\Theta}) = - i \langle \Psi (\bm{\Theta}) | \partial_{\bm{\Theta}}  \Psi (\bm{\Theta}) \rangle.
\end{eqnarray}
The Berry phase $\gamma$ is defined 
as a contour integral of the Berry connection as 
\begin{eqnarray}
\gamma =  \int_{L_1} d \bm{\Theta} \cdot \bm{A} (\bm{\Theta}) \hspace{3mm} ({\rm mod} \hspace{1mm}2\pi),
\end{eqnarray}
where $L_1$ is a path in a five-dimensional parameter space given as 
$L_1 = \bm{E}_0$ $\rightarrow$ $\bm{G}$ $\rightarrow$ $\bm{E}_1$ with 
$\bm{E}_0 = (0,0,0,0,0)$, $\bm{G} = \frac{1}{6} (2\pi,2\pi,2\pi,2\pi,2\pi)$, 
and $\bm{E}_1 = (2\pi,0,0,0,0)$. 
The Berry phase defined in this way is quantized in $\mathbb{Z}_6$
because there are five alternative paths $L_i$ ($i=2,3,4,5,6)$ which is equivalent to $L_1$
due to the six-fold symmetry of $H$, and the sum of the six Berry phases 
defined on each path $L_i$ is $0$ in modulo $2\pi$,
meaning that $\gamma = \frac{2\pi}{6} n$ ($n=0,1,2,3,4,5$);
see Refs.~\cite{Araki2019,Hatsugai2011} for the details. 

In Fig.~\ref{fig:4}, we plot $\gamma$ as a function of $t_0 / t_1$ (black open triangles).
Clearly, $\gamma = 0$ for $t_0/t_1 < 1$, 
while $\gamma= \pi$ $\left( = \frac{2\pi}{6} \cdot 3\right)$ for $t_0/t_1 >  1$.
This result indicates that the $\mathbb{Z}_6$ Berry phase is a topological invariant 
in that its value does not change as far as the energy gap remains open,
as we mentioned in the previous subsection. 
Consequently, the ground state for $t_0/t_1 > 1$ is adiabatically connected to that for the limiting 
case $t_0/t_1 \rightarrow \infty$, i.e., the hexamer state.

\subsection{$\mathbb{Z}_2$ Berry phase}
Similarly, the dimer state can be captured by
the $\mathbb{Z}_2$ Berry phase~\cite{Hatsugai2006} defined with respect to the local twist on a certain bond.
In the present model, the non-trivial $\mathbb{Z}_2$ Berry phase indicates that 
the ground state is adiabatically connected to the dimer state.
Importantly, the dimer state can be regarded as a HOTI too, since 
it hosts corner states under an appropriate choice of edges,
as demonstrated in Refs.~\cite{Liu2019,Lee2019,ZangenehNejad2019}.

The definition of the $\mathbb{Z}_2$ Berry phase is as follows:
We pick up one of the bonds which has the hopping $t_1$ (see an orange bold bond between sites A and B in Fig.~\ref{fig:1}).
Then, we again separate the Hamiltonian as $H = h_0^\prime + (H-h_0^\prime)$
where $h_0^\prime$ stands for the Hamiltonian on the bond we chose. 
We introduce the twisted Hamiltonian $H^\prime (\theta) = h_0^\prime(\theta) + (H-h_0^\prime)$, 
where 
\begin{eqnarray}
h^\prime_0(\theta) = t_1  
e^{i\theta}c^{\dagger}_{A} c_{B} + (\mathrm{h.c.}).
\end{eqnarray}
Then, for the ground state of $H^\prime (\theta)$, which we write as $|\Psi^\prime (\theta) \rangle$,
we define the Berry connection as 
\begin{eqnarray}
A^\prime(\theta) = - i \langle \Psi^\prime (\theta) | \partial_\theta  \Psi^\prime (\theta) \rangle,
\end{eqnarray}
and the corresponding Berry phase as 
\begin{eqnarray}
\gamma^\prime =  \int_{0}^{2\pi} d \theta \hspace{.5mm} A^\prime(\theta) \hspace{3mm} ({\rm mod} \hspace{1mm}2\pi).
\end{eqnarray}

In Fig.~\ref{fig:4}, we plot $\gamma^\prime$ as a function of $t_0 / t_1$ (blue open circles).
Contrary to $\gamma$, 
$\gamma^\prime = 0$ for $t_0/t_1 > 1$, 
while $\gamma = \pi$ for $t_0/t_1 <  1$.
Thus, similarly to the case of $\mathbb{Z}_6$ Berry phase, 
it is indicated that 
the ground state for $t_0/t_1 <  1$ is adiabatically connected to the dimer state. 

\section{Corner states on finite samples \label{sec:open}}
In this section, we show the energy spectra and wave functions under the open boundary conditions in both of two directions,
to demonstrate that the nontrivial Berry phase results in 
a topologically-protected boundary states at the corners, which is a defining feature of the HOTI phase.
Since the corner states for $t_0 < t_1$ have already been investigated in the previous works~\cite{Liu2019,Lee2019,ZangenehNejad2019},
we focus on the opposite case, i.e., $t_0 >  t_1$.

\begin{figure}[t]
\begin{center}
\includegraphics[width=.95\linewidth]{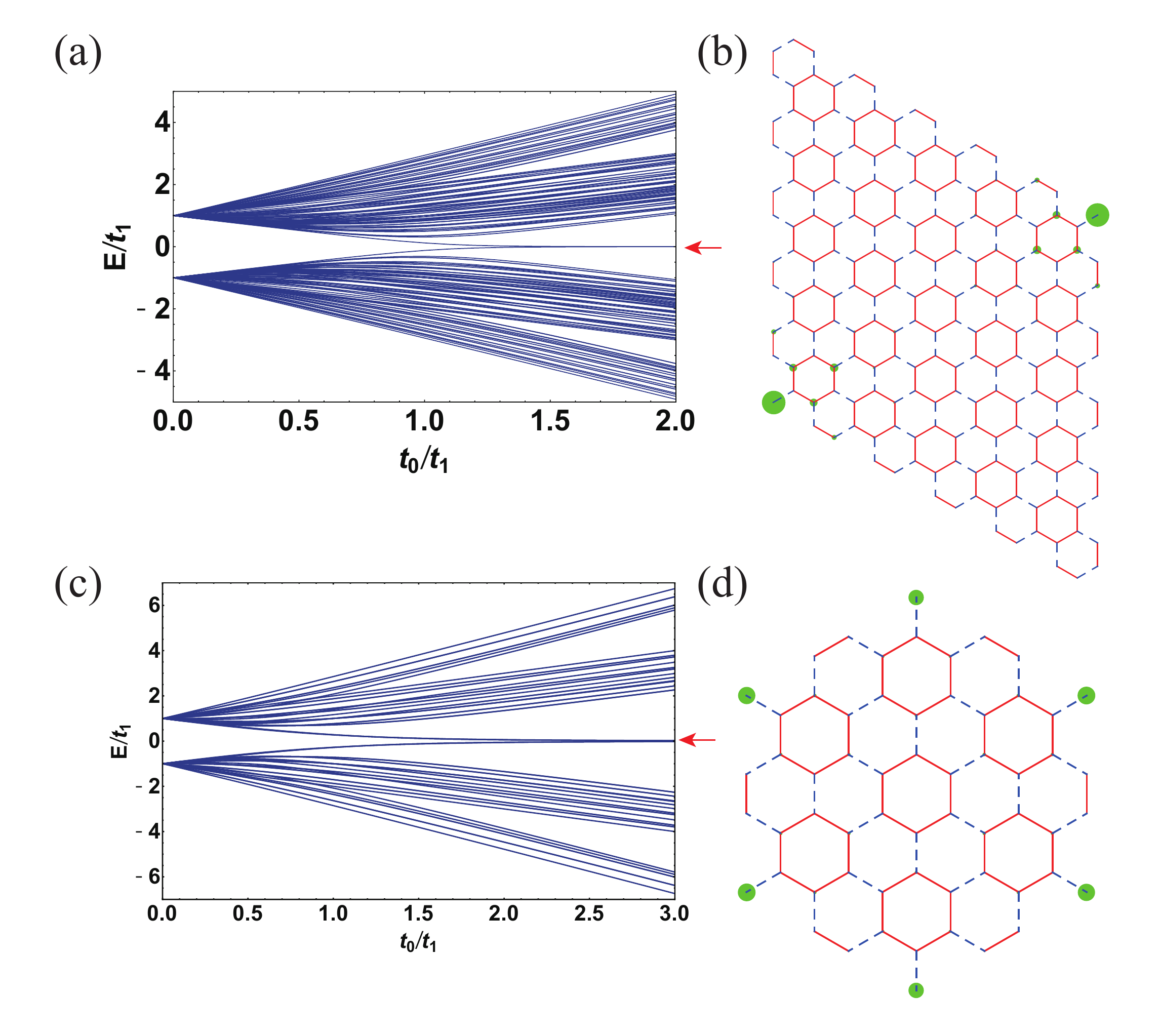}
\vspace{-10pt}
\caption{(Color online) (a) Energy spectrum 
of the present model
on a rhombus geometry under the open boundary condition
composed of 192 sites.
We see the in-gap states for $t_0 > t_1$ (see a red arrow).
(b) The real space distribution of the in-gap states for $t_0=1.5$, $t_1=1$.
An average is taken over two quasi-degenerate states.  The size of green dots denotes the amplitudes.
(c) Energy spectrum 
of the present model
on a hexagonal goemetry
under the open boundary condition composed of 60 sites.
We again see the in-gap states for $t_0 > t_1$ (see a red arrow).
(d) The real space distribution of the in-gap states for $t_0=2.5$, $t_1=1$.
An average is taken over six quasi-degenerate states.  
The size of green dots denotes the amplitudes.}
  \label{fig:5}
 \end{center}
\end{figure}

\begin{figure}[t]
\begin{center}
\includegraphics[width=.98\linewidth]{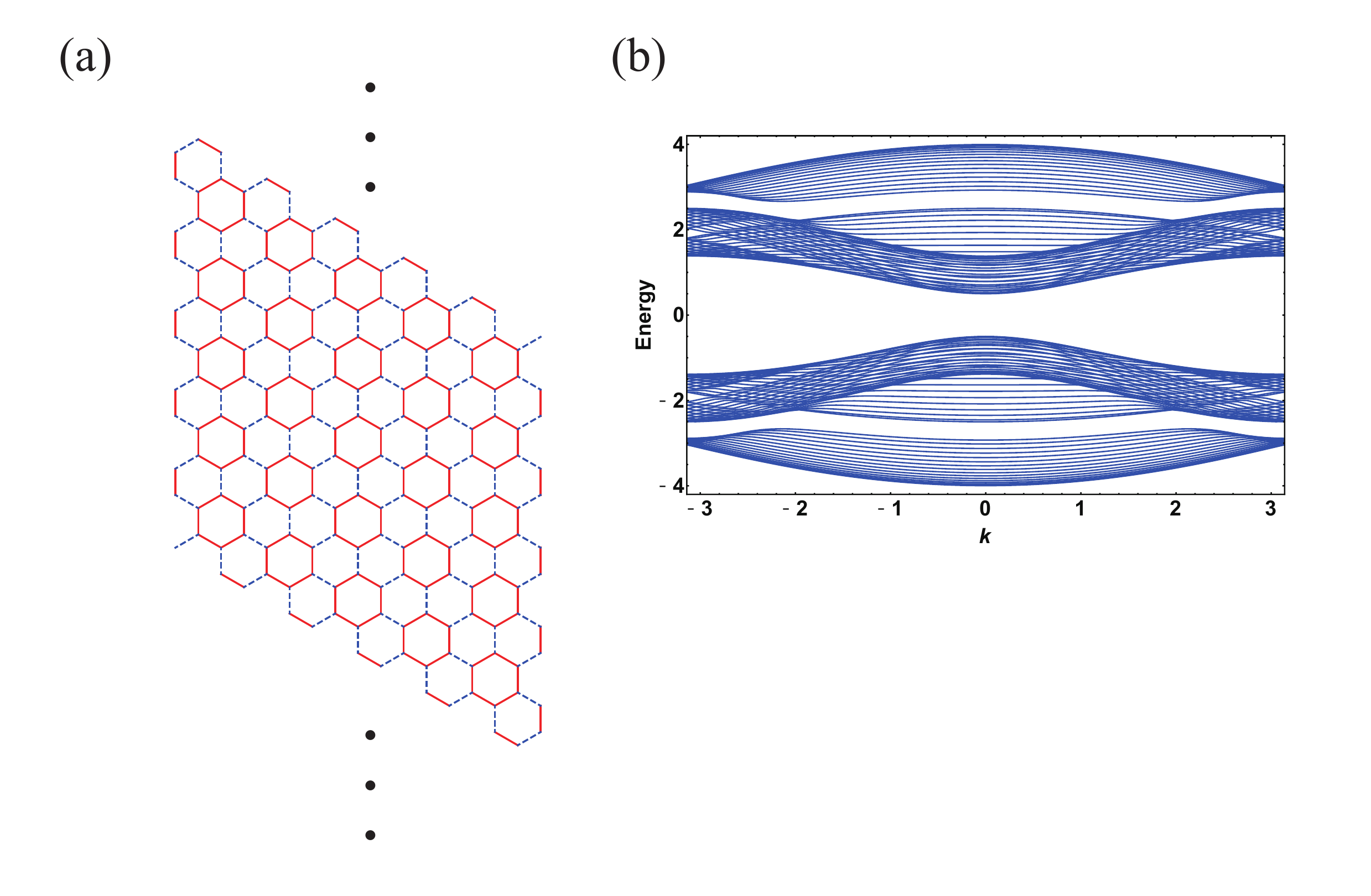}
\vspace{-10pt}
\caption{(Color online) (a) The present model on a cylinder. 
The dots mean that in we impose the periodic boundary condition in that direction. 
(b) The band structure for $t_0=1.5$ and $t_1 = 1$. }
  \label{fig:edge}
 \end{center}
\end{figure}

To gain the insight for the corner states in this system, 
it is useful to consider the ``irreducible cluster" limit.
The aforementioned picture of the hexamer state indicates that 
the localized state appears when the hexagons are cut off at the boundaries,
similarly to the end state of the SSH model which appears when the strong bond is cut off at the end.
This means that the boundary states in the present model crudely depend on the shape the boundary;
this coincides with the results in the previous work~\cite{Kariyado2017}.
In this paper, we consider the finite systems shown in Fig.~\ref{fig:5}(b), 
(a rhombus geometry) and Fig.~\ref{fig:5}(d), (a hexagonal geometry)
where 
the boundaries intersect hexagonal plaquettes at the edges and the corners.
In both of these geometries, the shape of edges is tailored as follows:
\begin{enumerate}
\item[(I)] Focusing only on the strong (i.e., solid) bonds, 
there are no open monomers, trimers, nor pentamers along the edges.
\item[(II)] There are isolated sites which are not connected to any of strong bonds only at 120-degree corners. 
\end{enumerate}
Clearly, the condition (II) is required so that the corner states appear. 
In addition, the condition (I) is required so that the edge states are gapped or the edge state does not exist.
Indeed, the edge state 
does not appear when we impose 
a periodic boundary condition in one of the directions for 
the geometry of Fig.~\ref{fig:5}(a), as shown in Fig.~\ref{fig:edge}.

In Figs.~\ref{fig:5}(a) 
and \ref{fig:5}(c),
we plot the energy spectra as a function of $t_0/t_1$. 
We do not see the in-gap state around zero energy for $t_0/t_1 < 1$,
whereas there exist in-gap states for $t_0/t_1 > 1$ 
which have quasi-two-fold degeneracy for Fig.~\ref{fig:5}(a) and quasi-six-fold degeneracy for Fig.~\ref{fig:5}(c).
Note that the energy is slightly deviated from zero due to the finite-size effect.  

Green circles in Fig.~\ref{fig:5}(b) [\ref{fig:5}(d)] represent the real-space distribution of the wave function, 
averaged over two [six] in-gap states. 
Clearly, it is localized at the 120-degree corners where the hexagonal plaquettes are cut off and isolated sites exist. 
Note that, if we remove one of the corner sites, 
the energy of one of the in-gap states is exactly pinned to zero due to the imbalance 
between two sublattices of the original honeycomb lattice~\cite{Brouwer2002}.

From these results, 
we conclude that the ground state is in the HOTI phase for $t_0/t_1 > 1$.

\section{$\mathbb{Z}_6$ Berry phase for $1/6$-filling \label{sec:1/6}}
So far, we have discussed the case of half-filling.
However, the non-trivial Berry phase which is a hallmark of adiabatic connection to the irreducible cluster states
is not inherent in the half-filled system. 
Indeed, the Berry phase is well-defined as far as the system is gapped,
and the non-trivial Berry phase indicates the adiabatic connection to the irreducible cluster states 
even for lower or higher fillings. 
To see this, in this section, we show the $\mathbb{Z}_6$ Berry phase for $1/6$-filling.

Although the $1/6$-filled system is gapped for an isolated hexamer [Fig.~\ref{fig:3}(c)], 
this is not the case for any $t_0 > t_1$.
As we mentioned in Sect.~\ref{sec:model}, 
the system is semimetallic for $1 < t_0/ t_1 \lesssim 1.32$.
We therefore focus on  the region $ t_0/ t_1 \gtrsim 1.32$, where the system is insulating and thus the $\mathbb{Z}_6$ Berry phase is well-defined.

\begin{figure}[b]
\begin{center}
\includegraphics[width=.95\linewidth]{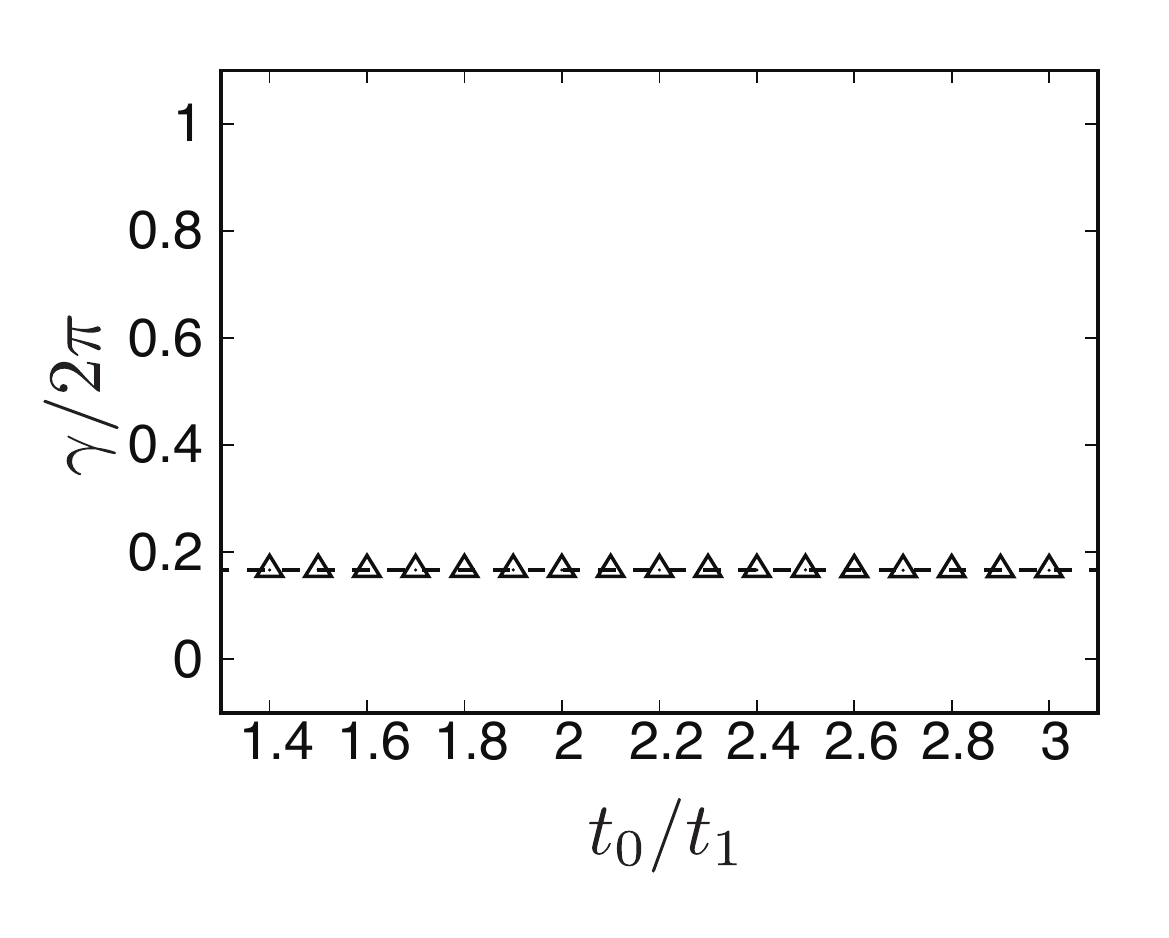}
\vspace{-10pt}
\caption{The $\mathbb{Z}_6$ Berry phase at $1/6$-filling.
Open triangles are the data for $t_0/t_1 = 0.1 n $ ($n= 14, \cdots 30$).
A dashed line denotes the Berry phase $\pi/3$. }
  \label{fig:6}
 \end{center}
\end{figure}

In Fig.~\ref{fig:6}, we plot the $\mathbb{Z}_6$ Berry phase for $1/6$-filling.
We see that the Berry phase is quantized as $\frac{\pi}{3}$ $\left(= \frac{2\pi}{6} \cdot 1\right)$,
indicating that the ground state is adiabatically connected to the hexamer state with $1/6$-filling.
Therefore, it is expected that the ground state for the $1/6$-filling is also a topological phase, 
accompanied by the edge or corner states depending on the shape of the boundaries.
We remark that the system with Kekul\'{e} modulation (i.e. $t_0 < t_1$) does not host 
the topological insulating phase for $1/6$-filling since it is gapless.

\section{Discussions and summary \label{sec:summary}}
In summary, 
we have shown that the HOTI phase is realized in
a honeycomb-lattice model
not only with Kekul\'{e} modulation but also with
anti-Kekul\'{e} modulation.
By means of the analyses on two different Berry phases,
namely the $\mathbb{Z}_6$ Berry phase and the $\mathbb{Z}_2$ Berry phase,
we have revealed that the ground state for the present model 
at half-filling is adiabatically connected the either of two irreducible cluster states,
namely, the hexamer state (for $t_0 > t_1$) and the dimer state (for $t_0 < t_1$).
For the former, the $\mathbb{Z}_6$ Berry phase becomes $\pi$, while for the latter, the $\mathbb{Z}_2$ Berry phase becomes $\pi$.
Since both of two states host topologically nontrivial structure 
in that they cannot be adiabatically connected to the atomic state,
the corner states appear under an appropriate choice of shape of the boundaries,
manifesting the existence of HOTI phase for both Kekul\'{e} modulation and 
anti-Kekul\'{e} modulation.
We have also elucidated that the $\mathbb{Z}_6$ Berry phase is $\frac{\pi}{3}$ for $1/6$-filling
with $t_0 \gtrsim 1.32 t_1$,
which indicates the adiabatic connection to the hexamer state at this filling as well.

So far, we have discussed the non-interacting fermion system, 
but we expect that our proposal of the HOTI phase 
opens up a way to search corner states in the various systems, 
such as correlated electron systems~\cite{Sorella2018},
photonic crystals~\cite{Wu2015,Yang2018,Li2018,Noh2018},
and surface electrons on metallic substrates with molecular scatterers~\cite{Freeney2019}.
Regarding the correlation effect, we remark that 
the Berry phase can be defined even in the presence of interactions~\cite{Araki2019,Kudo2019},
since it is defied for the many-body ground state.
This means that the Berry phase will be a powerful tool to identify the correlated HOTI phase.
For the spinless fermion model with weak inter-site repulsions whose strength does not exceed the band gap,
it was found in the BBH model that the many-body analog of the HOTI phase survives, which is identified by the Berry phase~\cite{Araki2019}.
As for the spinful fermion model with the Hubbard interaction,
the higher-order topological Mott insulating (HOTMI) phase~\cite{Kudo2019} which is characterized by 
gapped charge excitations and gapless spin excitations at the corners may be realized.  
Searching the realization of the HOTMI in this model will be an interesting future problem.

Before closing this paper, we address the robustness of the corner states against disorders, 
which is crucial for experimental observation of corner states.
In fact, it has been revealed in the breathing kagome model that the corner states of HOTIs are robust against 
disordered on-site potentials as far as the disorder strength is smaller than the energy gap~\cite{Araki2019_2}.
Hence, we expect that the corner states in the present model are robust against 
disordered on-site potentials as well.

\begin{acknowledgement}
We thank T. Kariyado for valuable comments. 
We also thank X. Hu for letting us know the relevant paper of Ref.~\cite{Noh2018}. 
This work is supported by the JSPS KAKENHI, Grant No. JP17H06138 and No. JP19J12315, MEXT, Japan.
\end{acknowledgement}

\end{document}